\documentclass[%
 aip,
 amsmath,amssymb,
preprint,
groupedaddress, %remove superscript on each author
]{revtex4-2}

\usepackage{graphicx}% Include figure files
\usepackage{dcolumn}% Align table columns on decimal point
\usepackage{bm}% bold math
\usepackage[utf8]{inputenc}
\usepackage[T1]{fontenc}
\usepackage{mathptmx}
\usepackage{amsmath}
\usepackage{xcolor}
\usepackage{soul}

\begin{document}

%\preprint{AIP/123-QED}

\title[Thermal modulation of ultrasonic wave]{Determination of acoustic nonlinearity parameters using thermal modulation of ultrasonic waves}
% Force line breaks with \\

\author{Hongbin Sun}
\email{hbsun1991@huskers.unl.edu}
\author{Jinying Zhu}%
 \email[Author to whom correspondence should be addressed:]{jyzhu@unl.edu.}
\affiliation{ 
Department of Civil and Environmental Engineering, University of Nebraska-Lincoln, 1110 S 67th St. Omaha, NE 68182, USA%\\This line break forced with \textbackslash\textbackslash
}%

\thanks{Published as: J. Zhu et al., Applied Physics Letters 116, 241901 (2020), DOI: 10.1063/5.0014975.}

\begin{abstract}
This study presents a test method and its theoretical framework to determine the acoustic nonlinearity parameters ($\alpha,\beta,\delta$) of material using thermal modulation of ultrasonic waves.  Temperature change induced thermal strain excites the nonlinear response of the material and modulates the ultrasonic wave propagating in it.  Experimental results showed a strong correlation between the relative wave velocity change and the temperature change. With a quadratic polynomial model, the acoustic nonlinearity parameters 
were obtained from the polynomial coefficients by curve fitting the experimental curves. Their effects on thermal-induced velocity change were discussed. The parameters $\alpha,\beta,\delta$ govern the hysteretic gap, average slope, and curvature of the correlation curve, respectively.  The proposed theory was validated on aluminum, steel, intact and damaged concrete samples. The obtained nonlinear parameters show reasonable agreements with values reported in the literature. Compared to other nonlinear acoustic methods using vibration or acoustic excitation, the thermal modulation method generates more uniform, slow changing, and larger strain field in the test sample. Employing thermal effect as the driving force for nonlinearity instead of an undesired influencing factor, this method can measure the absolute values of $\alpha,\beta,\delta$ with good accuracy using a simple ultrasonic test setup.  
\end{abstract}

\maketitle

% \begin{quotation}
% The ``lead paragraph'' is encapsulated with the \LaTeX\ 
% \verb+quotation+ environment and is formatted as a single paragraph before the first section heading. 
% (The \verb+quotation+ environment reverts to its usual meaning after the first sectioning command.) 
% Note that numbered references are allowed in the lead paragraph.
% %
% The lead paragraph will only be found in an article being prepared for the journal \textit{Chaos}.
% \end{quotation}

%\section{\label{sec:level1}Introduction}
Nonlinear acoustic/ultrasonic techniques show high sensitivity to microcracking damage, especially in complex materials. Following the phenomenological description of stress-strain hysteresis of rock by McCall and Guyer  \cite{mccall1994equation,Mccall1996hysteresis}, the elastic modulus of a complex material can be modeled as a strain and strain rate dependent variable and described by three nonlinear parameters $\beta$, $\delta$ and $\alpha$. The first two parameters represent the classical nonlinear perturbation parameters and $\alpha$ is the non-classical nonlinear parameter, which is introduced as a measure of material hysteresis. %Most nonlinear acoustic/ultrasonic test methods focus on measuring these nonlinear parameters and using them for nondestructive evaluation (NDE) of materials. 
Meurer et al. \cite{MEURER2002} proposed a general form of a nonlinear constitutive model, and they showed that the McCall and Guyer's model is a special case of this general model when only the leading order effect of $\alpha$ is considered. 

Acoustic nonlinearity of material may manifest as softening of modulus with increasing strain or higher harmonic generation.  Modulus softening can be measured as the change of resonance frequency or ultrasonic wave velocity at different strain levels. The nonlinear resonance acoustic spectroscopy (NRAS) method \cite{van1997quasi,van2000nonlinear2} measures the resonance frequency shift $\Delta f$ when the resonance mode in a sample is excited at different strain levels. The linear slope between $\Delta f$ and the strain change $\epsilon$ is defined as the relative nonlinear parameter $\alpha$. Dynamic acousto-elastic testing (DAET) method is a recently developed nonlinear ultrasonic method that measures the relative ultrasonic wave velocity change due to a low-frequency strain modulation  \cite{riviere2013pump,shokouhi2017dynamic}. Nonlinear parameters $\beta$, $\delta$, and $\alpha_{DAET}$ were derived from the correlation curves between the relative velocity change and strain.
Higher harmonic generation is a phenomenon wherein the ultrasonic waveform is distorted by the nonlinear response of the material, then higher harmonic waves are generated. Second harmonic generation (SHG)  \cite{breazeale1965ultrasonic,hikata1965dislocation} measures the amplitudes of the fundamental and the second harmonics to calculate the nonlinear parameter $\beta$. Because the absolute displacement amplitudes are challenging to measure, a relative value of $\beta$ is typically used for characterizing fatigue of metal materials \cite{kim2006experimental} and concrete damage \cite{kim2014air}.

Although these nonlinear ultrasonic testing methods show different degrees of success for extracting nonlinear parameters and characterizing material damage, some limitations hinder the application of these methods to practice. First, the absolute nonlinear parameters are not easy to measure. The NRAS method uses the acceleration instead of the strain to calculate an alternative nonlinear parameter $\alpha_f$.  In addition, NRAS is only applicable to small samples since it needs to excite the global vibration of the test sample. Second, many nonlinear ultrasonic methods use an impact or an actuator to excite nonlinear responses of the test sample. The excitation applied at a single point, or a small area creates a non-uniform strain field in a local area of the sample. The measurement results are affected by the strain distribution in the sample. 

In a recent work by the authors \cite{sun2019thermal}, we investigated ambient temperature induced ultrasonic wave velocity change in concrete samples with microcracking damage, and defined a thermal modulation coefficient to represent the sensitivity of wave velocity to the temperature effect. It is found that the thermal modulation coefficient increases with the damage level in concrete samples, which shows its potential for nondestructive evaluation of material. In this letter, we propose using the thermal modulation of nonlinear ultrasonic waves to determine the nonlinear parameters $\alpha,\beta,\delta$. Slow temperature change creates a relative uniform thermal strain field in the test sample. The thermal strain can be used as the driving force to excite nonlinear behaviors of a material and modulate ultrasonic waves propagating in the material. Based on the experimental results from the thermal modulation test, we built a theoretical model to determine the absolute values of nonlinear parameters.  The proposed theory was validated on classical nonlinear materials (metals) and non-classical nonlinear materials with hysteresis (concrete).

\begin{figure}[ht]
\centering
\includegraphics[width=6.5cm]{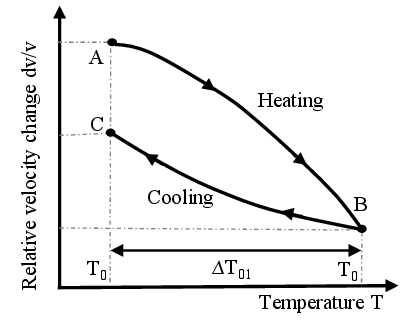}
\caption{ \label{fig:theory} Diagram for correlation between relative velocity change and temperature change}
\end{figure}

%\section{Thermal modulation of nonlinear ultrasonic wave}
 Temperature effects are often regarded as an undesired noise in most experimental studies. The dependence of elastic wave velocity on temperature has been studied by many researchers \cite{WEAVER2000,lu2005methodology,larose2009monitoring,sun2019thermal}.  Previous researchers \cite{WEAVER2000,lu2005methodology} have noticed a linear relationship between the relative velocity change $dv/v$ and the temperature change $\Delta T$ on metals. For materials with hysteresis, Fig.\ref{fig:theory} shows a diagram of $dv/v\sim T$ correlation, where a sample experienced a heating/cooling thermal cycle ($T_0\rightarrow T_1\rightarrow T_0$). The relative velocity change $dv/v$ decreases from point A to B during the heating and increases from B to C in the cooling. In this letter, through analytical derivations and experimental data analyses, we prove that the nonlinear acoustic parameters can be derived from the $dv/v\sim T$ correlations on metal materials and concrete. 

A quadratic polynomial is used to model the $dv/v$ vs. temperature relationship as:
\begin{equation}
dv/v=k_0+k_1\Delta T+k_2\Delta T^2,
\label{eq1}
\end{equation}
where $k_0$, $k_1$ and $k_2$ are the coefficients of the polynomial and $\Delta T$ is the temperature change relative to a reference temperature. These coefficients are obtained from curve fitting of the correlation curves in Fig.\ref{fig:theory}, and are denoted as the thermal modulation coefficients.

For nonlinear materials with hysteresis, the modulus can be described with a strain and strain rate related model as \cite{van1997quasi}: 
\begin{equation}
E(\varepsilon , \dot{\varepsilon})=E_0\left \{1-\beta\varepsilon-\delta \varepsilon ^2-\alpha \left [ \Delta\varepsilon+\varepsilon \text{sign}(\dot{\varepsilon}) \right ]    \right \},
\label{eq2}
\end{equation}
where $E_0$ is the linear modulus and $\beta$ and $\delta$ are the quadratic and cubic nonlinear parameters (classical nonlinear parameters). $\alpha$  describes the hysteretic behavior of non-classical nonlinear materials. $\varepsilon$ and $\dot{\varepsilon}$ are the strain and strain rate,  sign$(\dot{\varepsilon})=1$ if $\dot{\varepsilon}>0$ and sign$(\dot{\varepsilon})=-1$ if $\dot{\varepsilon}<0$. $\Delta\varepsilon$ is the maximum strain experienced in the previous loading cycles.

Since the velocity is related to $\sqrt{E}$, the relative velocity change is expressed as:
\begin{equation}
dv/v=\frac{1}{2}\frac{\Delta E}{E}=-\frac{1}{2}\left \{\beta\varepsilon+\delta \varepsilon ^2+\alpha \left [ \Delta\varepsilon+\varepsilon \text{sign}(\dot{\varepsilon}) \right ]    \right \}.
\label{eq3}
\end{equation}
Temperature change will generate thermal strain $\varepsilon$ in a material, which can be expressed as $\varepsilon=\alpha_T\Delta T $, where $\alpha_T$ is the thermal expansion coefficient.  If the sample experienced the temperature cycle as shown in Fig.\ref{fig:theory}, then $\Delta \varepsilon=\alpha_T\Delta T_{01}$ represents the maximum strain the sample experienced for both the heating and the cooling processes. 

For the heating process (A$\rightarrow$B), sign$(\dot{\varepsilon})=1$ and $T_0$ is used as the reference temperature. For the cooling process (B$\rightarrow$C), sign$(\dot{\varepsilon})=-1$ and $T_1$ is the reference temperature. Then equation (\ref{eq3}) can be written as below for the heating (+) and the cooling (-) processes 
\begin{equation}
   \begin{split}
      {dv/v}^{(\pm)}=-\frac{1}{2} \{\beta(\alpha_T\Delta T) +\delta (\alpha_T\Delta T)^2\\+ \alpha (\alpha_T\Delta T_{01}\pm\alpha_T\Delta T) \}.\\
      \end{split}
      \label{eq4}
\end{equation}
Equation (\ref{eq1}) for heating and cooling cycle can be rewritten as:
\begin{equation}
{dv/v}^{(\pm)}=k_0^\pm+k_1^\pm\Delta T+k_2^\pm\Delta T^2,
\label{eq5}
\end{equation}
where $k^\pm$ represents the coefficients for heating and cooling processes.

By comparing the first degree term $\Delta T$ in Eqs. (\ref{eq4} -\ref{eq5}), we can get the following solution for $k_1^\pm$:
\begin{equation}
\left\{\begin{matrix}
k_1^+=-\alpha_T (\beta+\alpha)/2,\\ 
k_1^-=-\alpha_T (\beta-\alpha)/2. 
\end{matrix}\right.
\label{eq6}
\end{equation}
The nonlinear parameters $\beta$ and $\alpha$ are solved as:
\begin{equation}
\left\{\begin{matrix}
\beta=-(k_1^-+k_1^+)/\alpha_T, \\ 
\alpha=-(k_1^+-k_1^-)/\alpha_T.
\end{matrix}\right.
\label{eq7}
\end{equation}
Equation (\ref{eq7}) indicates that for classical nonlinear materials without hysteresis ($\alpha=0$), the coefficients  $k_1^+$ and $k_1^-$ should be equal. This statement is validated in the experiments presented in this letter.

By comparing the second degree term $\Delta T^2$ in Eqs. (\ref{eq4} - \ref{eq5}), we obtain two solutions to the cubic nonlinear parameter $\delta$ from the heating and cooling processes, which is related to the curvatures of the correlation curves. 
\begin{equation}
\delta^\pm=-2k_2^\pm/\alpha_T^2. %\textrm{heating}\\ 
%\left\{\begin{matrix}
%\delta^+=-2k_2^+/\alpha_T^2\quad \textrm{heating}\\ %\delta^-=-2k_2^-/\alpha_T^2 \quad \textrm{cooling}.
%\end{matrix}\right.
\label{eq8}
\end{equation}

%\section{Experimental results and discussions}
To perform the thermal modulation test, the samples were placed in an environmental chamber with a controllable temperature changing rate. The sample experienced a slow temperature changing rate ($1 ^\circ$C/hr) to minimize the temperature gradient. Two 2.25 MHz ultrasonic transducers {(Olympus A106S)} were installed on the two opposite surfaces of the metal samples as the transmitter and receiver. %(see Fig.\ref{fig:setup})%. 
For concrete samples, two 15 mm diameter {PZT discs (STEMINC SMD15T21R111WL) were used to excite and receive ultrasonic waves around the frequency of 150 kHz}. Ultrasonic signals were acquired with a digital oscilloscope (PICO 4224), and the temperature was monitored using a Type T thermocouple and a data logger. 

For common structural materials (aluminum, steel, and concrete), the relative velocity change with temperature are in the order of $10^{-4}/^{\circ}$C. In order to accurately measure such small velocity changes, we used the the coda wave interferometry (CWI) method \cite{lobkis2003coda}. If two signals differ only by a dilation, which is typically true for temperature-induced signal distortion, we can determine the dilation by stretching the disturbed signal and obtaining the stretching factor when the cross-correlation between the signals reaches maximal. In the experiments, we were able to measure the relative velocity change with a precision of $10^{-6}$.

The dilation $\delta_t=dt/t$ represents the relative time delay $dt$ at the time window $t$ of a signal.  In the thermal modulation test, two effects contribute to the relative time delay $\delta_t$: thermal expansion of the sample, and temperature dependence of the wave velocity \cite{lobkis2003coda,lu2005methodology}. The relative velocity change will be calculated as $dv/v=- (\delta_t-\alpha_T\Delta T)$.  If the thermal strain is negligible ($\epsilon=\alpha_T\Delta T \ll dv/v$), the relative velocity change can be calculated as $dv/v=-dt/t=\delta_t$.

\begin{figure}[!htb]
\centering
\includegraphics[width=7.5cm]{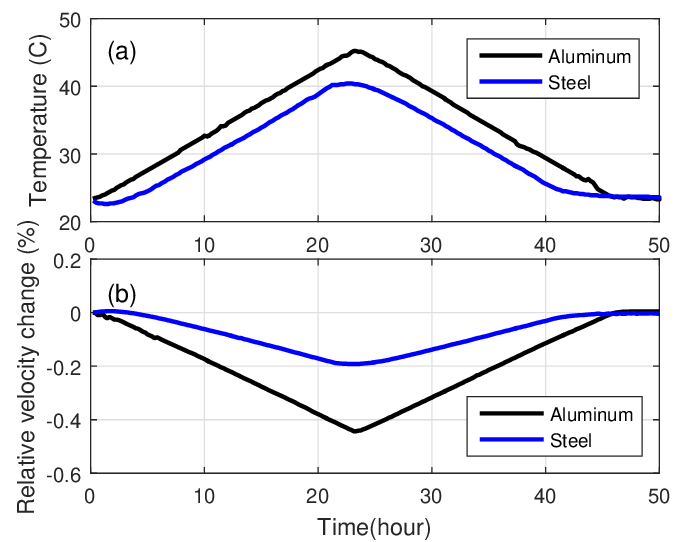}
\caption{ \label{fig:ALhistory} Thermal modulation test results of aluminum and steel: (a) temperature history, and (b) relative velocity change history}
\end{figure}

We first tested metal materials (aluminum and steel) using the thermal modulation method. Figure \ref{fig:ALhistory} shows the thermal modulation results of an aluminum 6061 block (5 cm$\times$5 cm$\times$18 cm) and a stainless steel 304 block (3 cm$\times$6 cm$\times$30 cm) with the temperature histories and relative velocity change histories. The two samples were first heated to a specific temperature and then cooled back to room temperature. The temperature changing rate was about 0.96$^\circ$C/hour, which was close to the designed temperature rate. The relative velocity change in Fig. \ref{fig:ALhistory}(b) shows an opposite trend to the temperature change history.

\begin{figure}[!htb]
\centering
\includegraphics[width=7.5cm]{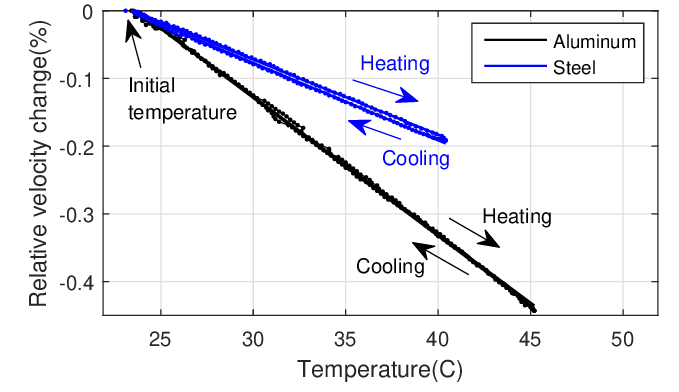}
\caption{ \label{fig:Al_Steel} Thermal modulation test results on aluminum 6061 and stainless steel sample. \small\textit{[Note for arXiv preprint: The slight hysteresis in the steel results is due to minor temperature measurement error rather than material response.]}}
\end{figure}

\begin{table}[!htb]
\caption{\label{tab1}Nonlinear parameters for Aluminum and Steel}
\begin{ruledtabular}
\begin{tabular}{l|c|c}
 Sample&  Aluminum 6061  & Stainless steel 304\\
\hline
$\alpha_T$ &$23.0\times10^{-6}$  & $17.3\times10^{-6}$\\
$k_1^+$, $k_2^+$         &-$1.77\times10^{-4}, -9.3\times10^{-8}$ & $-0.93\times10^{-4}, -5.8\times10^{-7}$\\
$k_1^-$, $k_2^-$         &-$1.77\times10^{-4}, -1.1\times10^{-7}$ & $-0.93\times10^{-4}, +6.0\times10^{-7}$\\
$\beta$ &15.4 &\textbf10.7\\
%$\beta$\cite{dace1992measurement,yost1993effects,ostrovsky2001dynamic,torello2017determina%tion} &4\textasciitilde12 &2\textasciitilde4.5\\
$\delta^+$,$\delta^-$ &351, 415 & 3875, -4009\\
$\alpha$ &0 &0\\
\end{tabular}
\end{ruledtabular}
\end{table}
Figure \ref{fig:Al_Steel} presents the $dv/v\sim T$ curves for the aluminum and the steel samples. The extracted thermal modulation coefficients are summarized in Table \ref{tab1}. 
The steel sample has a smaller absolute value of $k_1^\pm$ (slope) than the aluminum sample. Both correlation curves show very good linearity, which indicates small values of $|k_2^\pm|$ for aluminum and steel.

The nonlinear parameters $\alpha,\beta,\delta$ can be calculated from Eqs. (\ref{eq7},\ref{eq8}) by using the thermal expansion coefficients $\alpha_T$=23 $\mu \varepsilon/^\circ$C for aluminum and $\alpha_T$=17.3 $\mu \varepsilon/^\circ$C for steel. The results are also given in Table \ref{tab1}. From the thermal modulation tests, we obtained $\beta= 15.4$ for aluminum and $\beta=10.7$ for steel. The absolute value of $\beta$ reported in literature varies from 4 to 12 for aluminum alloys and 2 to 4.5 for stainless steel \cite{dace1992measurement,yost1993effects,ostrovsky2001dynamic,torello2017determination}. The $\beta$ values obtained from the thermal modulation test are slightly larger than the reported values in literature for both aluminum and steel. A possible explanation is the differences in strain measurement and strain range. In the thermal modulation test, the samples had a very slow strain changing rate, relatively large strain level, and uniform strain distribution, while most other nonlinear acoustic methods (NRAS, DAET, SHG) generate dynamic strain at very low strain levels.

For both the aluminum and the steel curves, the heating and cooling slopes are equal $k_1^+=k_1^-$, which gives the non-classical parameters $\alpha=0$. Although both curves show good linearity, the $\delta^\pm$ values for the steel are almost 10 times higher than that of the aluminum sample. The steel sample also shows a narrow hysteresis area. Just like the difference between {$k_1^+$ and $k_1^-$} gives the hysteresis parameter $\alpha$, the authors believe that the difference between $\delta^+$ and $\delta^-$ may represent a higher order hysteretic response ($\sim\epsilon^2$). This higher order effect of hysteresis has been predicted by Meurer et al. \cite{MEURER2002} in their general nonlinear model, while Eq. (\ref{eq3}) only includes the lower order term ($\sim\epsilon$). Since samples will experience large thermal strains ($10^{-4}\sim10^{-3} $) in the thermal modulation test, the higher order effects of $\alpha$ may need to be considered in future studies.

\begin{figure}[!htb]
\centering
\includegraphics[width=8cm]{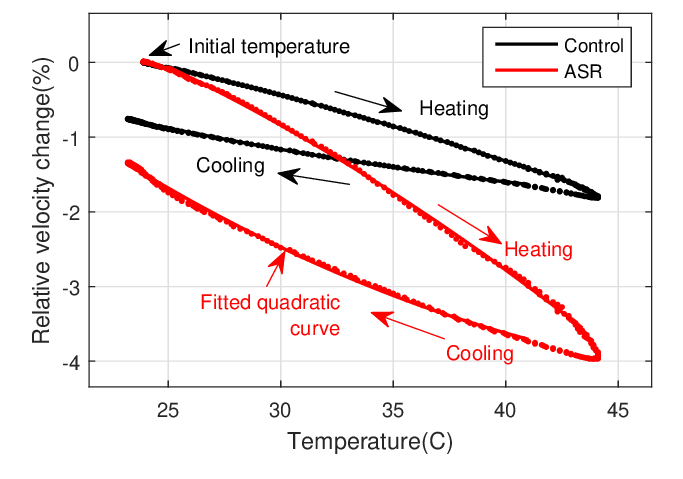}
\caption{ \label{fig:concrete} Thermal modulation test results of concrete}
\end{figure}

Concrete is a typical non-classical nonlinear material. We tested two concrete cylinders (10 cm diameter$\times$20 cm height), with one control sample cast with normal concrete mix, and another sample with microcracking damage induced by alkali-silica reaction (ASR). ASR is a chemical reaction in concrete that occurs between alkali hydroxides in hydrated cement and reactive silica in certain types of aggregates. ASR produces an expansive gel, which will eventually cause microcracks and degradation of concrete. The mix design and curing condition of the ASR sample were described in a previous work by the authors \cite{sun2019ultrasonic}. At the test time, the ASR sample had moderate damage (0.1\% expansion) with minor surface cracks while the control sample was in intact condition. For the thermal modulation test, the two concrete samples were first heated from 23.2$^\circ$C to 44.1$^\circ$C then cooled down to 23.2$^\circ$C. The correlation curves between the relative velocity change and temperature are shown in Fig.\ref{fig:concrete} for the two concrete samples. 

The thermal modulation coefficients $k_1^+, k_2^+$ and $k_1^-, k_2^-$ were obtained from curve fitting and are summarized in Table \ref{tab2}. All fittings have the goodness-of-fit $R^2$ larger than 0.99. For both the control and the ASR samples, there are $|k_1^+|>|k_1^-|$, which means that the relative velocity change is more sensitive to temperature change in the heating process than in the cooling process. The ASR sample has larger coefficients $|k_1^\pm|$ and $|k_2^\pm|$ than the control sample, which indicates that microcracking damage would increase the sensitivity of velocity change to temperature.  In the previous work by the authors \cite{sun2019thermal}, we also found the $|k_1^-|$ value increased with the damage level in concrete samples.  

The thermal expansion coefficients were measured for both the control and ASR samples. The results were very close, so that we used $\alpha_T$=10 $\mu\epsilon/^\circ$C for both samples. The nonlinear parameters $\alpha$, $\beta$, and $\delta$ were calculated using Eqs. (\ref{eq7}) and (\ref{eq8}). The absolute value of  $\beta$ is 100 for the control sample and 221 for the ASR sample. For the control sample, the result agrees with the literature reported values in the range of 40 to 157 for concrete \cite{larose2009monitoring,payan2009}.  The ASR sample shows a higher $\beta$ value than the control, which indicates that microcracking damage increases the quadratic nonlinearity. 

For both samples, $|\delta^+|$ and $|\delta^-|$ in the heating and the cooling processes are very close, and they are much larger than $|\delta^\pm|$ of the aluminum and the steel samples. The ASR sample has larger $|\delta^\pm|$ than the control sample, but the values are still on the same order of magnitude $10^5$, and are close to the reported value of $10^6$ for rock\cite{Guyer1999}. 

The ASR sample also shows a larger value of $\alpha=57$ than the control sample $\alpha=36$, which suggests that the damage in concrete increases the hysteresis. In Eqs. (\ref{eq4}-\ref{eq5}), $\alpha$ can also be solved from the constant term $|k_0^\pm|$. In a closed thermal cycle with the same starting and ending temperature, $\alpha$ is related to the gap in the $dv/v$ curves at the starting/ending points, i.e.  $\alpha=\Delta{(dv/v)}/(\alpha_T\Delta{T_{01}})$.  The $\alpha$ value calculated from the gap is 40 for the control sample and 73 for the ASR sample.

\begin{table}
\caption{\label{tab2}Nonlinear parameters for concrete samples}
\begin{ruledtabular}
\begin{tabular}{l|c|c}
 Sample&  Control  & ASR sample\\
\hline
$\alpha_T$ ($/^\circ$C) &$10.0\times10^{-6}$   & $10.0\times10^{-6}$\\
$k_1^+$, $k_2^+$      &-$6.8\times10^{-4},-9.5\times10^{-6}$ & $-13.9\times10^{-4},-2.8\times10^{-5}$\\
$k_1^-$,$k_2^-$      &-$3.2\times10^{-4},+8.9\times10^{-6}$ & $-8.2\times10^{-4},+2.9\times10^{-5}$\\
$\beta$ &100 &221\\
%$\beta$\footnotemark[1]  &40\textasciitilde157 &N/A \\
$\delta^+$,$\delta^-$& $1.9\times10^{5},-1.8\times10^{5}$&$5.6\times10^{5},-5.8\times10^{5}$\\
$\alpha$ & 36&57\\
\end{tabular}
\end{ruledtabular}
%\footnotetext[1]{Reported in literature}\\
\end{table}

The effects of nonlinear parameters on the $dv/v\sim T$ curve are summarized.  In Eq. (\ref{eq7}), $\beta$ and $\alpha$ were derived from the first degree coefficients $k_1^+$ and $k_1^-$, where $\beta$ is related to the average slope, and $\alpha$ associates to the difference between the heating/cooling slopes.  Therefore, $\beta$ gives the sensitivity of the relative velocity change to temperature change,  and $\alpha$ describes the material hysteresis. In addition, $\alpha$ contributes to the gap in $dv/v$ between points A and C in Fig.\ref{fig:theory}. For materials with very small hysteresis, the velocity will return to the original value after a closed thermal cycle. For mesoscopic materials (concrete, rock), the velocity may not go back to the original value, and this phenomenon has been reported by several researchers \cite{niederleithinger2013,guyer2009nonlinear}. The parameter $\delta$ is linked to the curvature of the correlation curves. The aluminum and steel samples have very small $\delta$ values, and their $dv/v\sim T$ curves show high linearity. For concrete samples, the correlation curves have similar curvatures during the heating and cooling processes, but with opposite signs. The ASR sample has a larger curvature than of the control sample, which indicates a larger $\delta$ value. The difference between $\delta^+$ and $\delta^-$ may be related to a higher order hysteresis behavior predicted by Meurer et al. \cite{MEURER2002}. In summary, for $dv/v\sim T$ curves from a thermal modulation test, hysteresis ($\alpha$) determines the gap, the difference between the heating/cooling slopes and curvatures; $\beta$ affects the average slope; $\delta$ controls the curvature of curves. Similar analyses were also given by Shokouhi et al. \cite{shokouhi2017dynamic} using the DAET method. Comparing the three parameters for the control and ASR concrete samples, we also conclude that microcracks and cracks increase both classical nonlinearity ($\beta$ and $\delta$) and non-classical nonlinearity-hysteresis ($\alpha$).      

%\section{Conclusions}
In this letter, we present a nonlinear ultrasonic test method based on thermal modulation to determine the absolute values of acoustic nonlinearity parameters of materials. Using a quadratic model, we derived the nonlinear parameters $\alpha, \beta, \delta$ from the $dv/v\sim T$ curve and determined their values by curve fitting of experimental data.  Compared to other nonlinear acoustic methods based on vibration or acoustic excitation, the thermal modulation method generates nonlinear behaviors of materials with a larger, relatively uniform, and slowly changing strain field. Therefore the relative velocity change and the thermal strain can be readily measured with reasonable accuracy without using complicated test systems and calibration procedures.  The experimental results showed reasonable agreements with literature reported values, which further validates that the nonlinear response is induced by material strain, either mechanical strain or thermal strain. Ongoing studies are focused on the quantitative correlation between the nonlinear parameters and ASR induced microcracking damage in concrete, which could provide a new nondestructive  approach to  evaluate material damage in the laboratory and in-situ tests.

This research is supported by the U.S. Department of Energy - Nuclear Energy University Program (NEUP) under the contract DE-NE0008544. 

\section {DATA AVAILABILITY}
The data that support the findings of this study are available
from the corresponding author upon reasonable request.

%\nocite{*}
 \bibliographystyle{aipnum4-1} 
\bibliography{aipsamp}% Produces the bibliography via BibTeX.

\end{document}